\documentclass[conference]{IEEEtran}
\IEEEoverridecommandlockouts
\usepackage{subcaption}

\usepackage{caption}
\usepackage{amsmath}
\usepackage{epsfig}
\usepackage{bm}
\usepackage{slashbox}
\usepackage{algorithmic}
\usepackage{latexsym}
\usepackage{amsmath}
\usepackage{amssymb}
\usepackage{epsfig}
\usepackage{amsthm}
\usepackage{graphicx}
\usepackage[ruled,vlined]{algorithm2e}
\usepackage{epstopdf}
\newtheorem{theorem}{Theorem}

\begin{document}
\title{\huge Social Network Enhanced Device-to-Device Communication Underlaying Cellular Networks}\vspace{-5cm}
\author{Yanru Zhang, \thanks{Y. Zhang (yzhang82@uh.edu) and Z. Han (zhan2@uh.edu) are with the Department of Electrical and Computer Engineering, University of Houston, Houston, Texas 77004.}
Lingyang Song,\thanks{L. Song (lingyang.song@pku.edu.cn) is with the School of Electrical Engineering and Computer Science, Peking University, Beijing, China, 100871.}
Walid Saad,\thanks{W. Saad (walid@miami.edu) is with the Electrical and Computer Engineering Department, University of Miami, Coral Gables, Florida 33146.}
Zaher Dawy,\thanks{Z. Dawy (zg03@aub.edu.lb) is with the Electrical and Computer Engineering Department, American University of Beirut, Beirut, Lebanon.}
and Zhu Han
\thanks{This work was made possible by NPRP grant $\sharp4-347-2-127$ from the Qatar National Research Fund (a member of Qatar Foundation). The statements made herein are solely the responsibility of the authors.}} \medskip
\maketitle

\begin{abstract}
Device-to-device (D2D) communication has seen as a major technology to overcome the imminent wireless capacity crunch and to enable new application services. In this paper, we propose a social-aware approach for optimizing D2D communication by exploiting two layers: the social network and the physical wireless layers. First we formulate the physical layer D2D network according to users' encounter histories. Subsequently, we propose an approach, based on the so-called Indian Buffet Process, so as to model the distribution of contents in users' online social networks. Given the social relations collected by the Evolved Node B (eNB), we jointly optimize the traffic offloading process in D2D communication. In addition, we give the Chernoff bound and approximated cumulative distribution function (CDF) of the offloaded traffic. In the simulation, we proved the effectiveness of the bound and CDF. The numerical results based on real traces show that the proposed approach offload the traffic of eNB's successfully.
\end{abstract}

\vspace{-0.15cm}
\section{Introduction}
\vspace{-0.2cm}
The recent proliferation of smartphones and tablets has been seen as a key enabler for anywhere, anytime wireless communications. The rise of online services, such as Facebook and YouTube, significantly increases the frequency of users' online activities. Due to this continuously increasing demand for wireless access, a tremendous amount of data is circulating over today's wireless networks. This increase in demand is straining current cellular systems, thus requiring novel approaches for network design.

In order to cope with this wireless capacity crunch, device-to-device (D2D) communication underlaid on cellular systems, has recently emerged as a promising technique that can significantly boost the performance of wireless networks \cite{Xu.2012}. In D2D communication, user equipments (UEs) transmit data signals to each other over a direct link instead of through the wireless infrastructure, i.e., the cellular network's Evolved Node Bs (eNBs). The key idea is to allow direct D2D communication over the licensed band and under the control of the cellular system's operator \cite{Xu.2012}.

Recent studies have shown that the majority of traffic in cellular systems consists of downloading contents such as videos or mobile applications. Usually, popular contents, such as certain YouTube videos, are requested more frequently than others. As a result, eNBs often end up serving different mobile users with the same contents using multiple duplicate transmissions. In this case, following the eNB's first transmission of the content, such content is now locally accessible to others in the same area, if UEs' resource blocks (RBs) can be shared with others. Newly arriving users that are within the transmission distance can receive the ``old" contents directly from those users through D2D communication. Here, the eNB only serves users that request ``new" content, which has never been downloaded. Through this D2D communication, we can reduce considerable redundant requests to eNB, so that the traffic burden of eNB can be released.

Our main contribution is to propose a novel approach to D2D communication, which allows to exploit the social network characteristics so as to reduce the load on the cellular system. To achieve this goal, first, we propose an approach to establish a D2D subnetwork to maintain the data transmission successfully. As a D2D subnetwork is composed by individual users, the connectivity among users can be intermittent. However, the social relations in real world tend to be stable over time. Such social ties can be utilized to achieve efficient data transmission in the D2D subnetwork. We name this social relation assisted data transmission wireless network by offline social network (OffSN).

Second, we assess the amount of traffic that can be offloaded, i.e., with which probability can the requested contents be served locally. To analyze this problem, we study the probability that a certain content is selected. This probability is affected by both external (influence from media or friends) and internal (user's own interests) factors. While users' interests are difficult to predict, the external influence which is based on users' selections can be easily estimated. To this end, we define an online social network (OnSN) that encompasses users within the OffSN, which reflect users' online social ties and influence to each other.

In this paper, We adopt practical metrics to establish OffSN, and the Indian Buffet Process to model the influence in OnSN. Then we can get solutions for the previous two problems. Latter we will integrate OffSN and OnSN to carry out the traffic offloading algorithm. Further more, in order to evaluate the performance of our algorithm, we will set up the Chernoff bound of the number of old contents user selects. To make the analysis more accurate, we also derive the approximated probability mass function (pmf) and cumulative distribution function (CDF) of it. From the numerical results, we show that under certain circumstances, our algorithm can reduce a considerable amount of of eNB's traffic. Our simulations based on the real traces also proved our analysis for the traffic offloading performance.

\vspace{-0.2cm}
\section{System Model}
\begin{figure}[!t]
\vspace{-0.45cm}
  \begin{center}
    \includegraphics[width=0.7\columnwidth,height=0.2\textwidth]{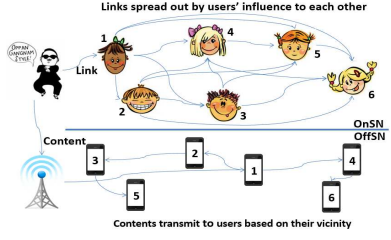}
    \vspace{-0.18cm}
    \caption{\label{fig:OnSN and OffSN}Information dissemination in both OnSN and OffSN.}
  \end{center}\vspace{-0.87cm}
\end{figure}

\vspace{-0.15cm}
Consider a cellular network with one eNB and multiple users. In this system, two network layers exist over which information is disseminated. The first layer is the OnSN, the platform over which users acquire the links of contents from other users. Once a link is accessed, the data package of contents must be transmitted to the UE through the actual physical network. Taking advantage of the social ties, the OffSN represents the physical layer network in which the requested contents of links to transmit. An illustration of this proposed model is shown in Fig.\ref{fig:OnSN and OffSN}. Each active user in the OnSN corresponds to the UE in the OffSN. In the OffSN, the first request of the content is served by the eNB. Subsequent users can thus be served by previous users who hold the content, if they are within the D2D communication distance.
\vspace{-0.55cm}
\subsection{Offline Social Network Model}
\vspace{-0.1cm}
In the area covered by an eNB, the density of users in public areas such as office buildings and commercial sites, is much higher than that in other locations such as sideways and open fields. Indeed, the majority of the data transmissions occurs in those fixed places. In such high density locations, forming D2D networks as an OffSN becomes a natural process. Thus, we can distinguish two types of areas: highly dense areas such as office buildings, and ``white" areas such as open fields. In the former, we assume that D2D networks are formed based on users' social relations. While in the latter, due to the low density, the users are served directly by the eNB. The OffSN is a reflection of the local users' social ties. Proper metrics need to be adopted to depict the degree of the connections among users. In \cite{Li.IEEE Conf2009}, the authors identify that human mobility shows a very high degree of temporal and spatial regularity, and that each individual returns to a few highly frequented locations with a significant probability. Thus, such social ties lead to higher probabilities to transmit data among users.

The contact duration distribution between two users is assumed to be a continuous distribution, which has a positive value for all real values greater than zero. In addition, users' encounter duration usually centers around a mean value. So we can adopt a $\Gamma(k,\theta)$ distribution which is widely used in modeling the call durations \cite{Guo.WiCom 2007} to model the call duration between two users. To find the value for the two parameter $k$ and $\theta$, we need the mean and variance of the contact duration.

\begin{figure}[!t]
\vspace{-0.45cm}
  \begin{center}
    \includegraphics[width=0.60\columnwidth,height=0.116\textwidth]{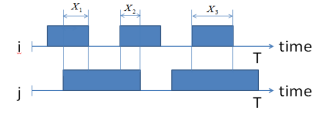}
    \vspace{-0.26cm}
    \caption{\label{fig:encounter history}Contact history between UE i and UE j.}
  \end{center}\vspace{-0.87cm}
\end{figure}

As shown in Fig. \ref{fig:encounter history}, given the contact duration $X_n$ and the number of encounters $N_{i,j}$ between UE $i$ and UE $j$ corresponding to user $i$ and user $j$ in the OnSN, an estimate of the expected contact duration length $M_{i,j}$ and variance $I_{i,j}$ can be given by:
\vspace{-0.3cm}
\begin{equation}
M_{i,j} = \frac {\sum_n X_n}{N_{i,j}},
\end{equation}\vspace{-0.3cm}
\begin{equation}\vspace{-0.3cm}
I_{i,j} = \frac {\sum_n (X_n - M_{i,j})^2} {N_{i,j}}.
\end{equation}\vspace{-0.23cm}

Given the mean and variance of the contact duration, we can derive the contact duration distribution: $X\sim \Gamma(k,\theta)= \Gamma(\frac{M_{i,j}^2}{I_{i,j}}, \frac{I_{i,j}}{M_{i,j}})$. Thus, the probability density function (PDF) of contact duration will be given by:
\begin{equation}\vspace{-0.25cm}
f(x; k, \theta) = \frac{1}{\theta^k}\frac{1}{\Gamma(k)}x^{k-1}e^{-\frac{x}{\theta}},
\end{equation}\vspace{-0.04cm}
where $\Gamma(k)=\int_0^\infty t^{k-1}e^{-t}dt$. Then, we can calculate the probability of the contact durations that are qualified for data transmission. If the contact duration is not sufficient to complete a data package transmission, the communication session cannot be carried out successfully. We adopt a closeness metric, $w_{i,j}$ to represent the probability of establishing a successful communication period between UE $i$ and UE $j$, which ranges from $0$ to $1$. The qualified contact duration is the complementary of the disqualified communication duration probability, so $w_{i,j}$ can be given by:
\begin{equation}\vspace{-0.25cm}
w_{i,j}=1-\int_0^{X_{min}}f(u;k,\theta)du=1-\frac{\gamma(k,\frac{X_{min}}{\theta})}{\Gamma(k)},
\end{equation}\vspace{-0.03cm}
where $X_{min}$ is the minimal contact duration required to successfully transmit one content data package, $\gamma(k,\frac{X_{min}}{\theta})=\int_0^{X_{min}} t^{k-1}e^{-t}dt$ is the lower incomplete Gamma function.

Hence, we can use the closeness metric $w_{i,j}$ to describe the communication probability between two UEs, which can also be seen as the weight of the link between UE $i$ and UE $j$. Then, a threshold $w_T$ can be defined to filter the boundary between different OffSNs and ``white" areas. To cluster users based on metrics such as closeness, we can adopt algorithms such as those used in social networks as in \cite{Li.IEEE Conf2009}. Then, with a properly chosen $w_T$, each pair in the OffSN will have a strong direct neighboring relationship.

\vspace{-0.1cm}
\subsection{Online Social Network Model}
\vspace{-0.1cm}

OnSN is the platform for content links to disseminate. We define the number of users in the OnSN as $N$ which, in turn, corresponds to $N$ UEs in the OffSN. The total number of available contents in the OnSN is denoted by $K$, $K = K_h + K_0$. Given the large volume of content available online, $K \rightarrow \infty$. $K_h$ represents the set of contents that have viewing histories and $K_0$ is the set of contents that do not have any. We adopt the Indian Buffet Process (IBP) \cite{Griffiths.ML 2011} model which serves as a powerful tool for getting the content popularity distribution and predicting users' selections.

The IBP is a stochastic process in which each diner samples from an infinite selection of dishes on offer at a buffet. The first customer will select its preferred dishes according to a Poisson distribution with parameter $\alpha$. Since all dishes are new to this customer, no external information exists so as to influence the selection. However, once the first customer completes the selection, the following customers will have some prior information about those dishes based on the first customer'’s feedback. Customers learn from the previous selections to update their beliefs on the dishes and the probabilities with which they will choose the dishes. The behavior of content selection behavior in OnSN is analogous to the dish selection in an IBP. If we view our OnSN as an Indian buffet, the online contents as dishes, and the users as customers, we can interpret the contents spreading process online by an IBP. So the probability distribution can be implemented from the IBP directly.

\begin{figure}[!t]
\vspace{-0.45cm}
  \begin{center}
    \includegraphics[width=0.71\columnwidth,height=0.19\textwidth]{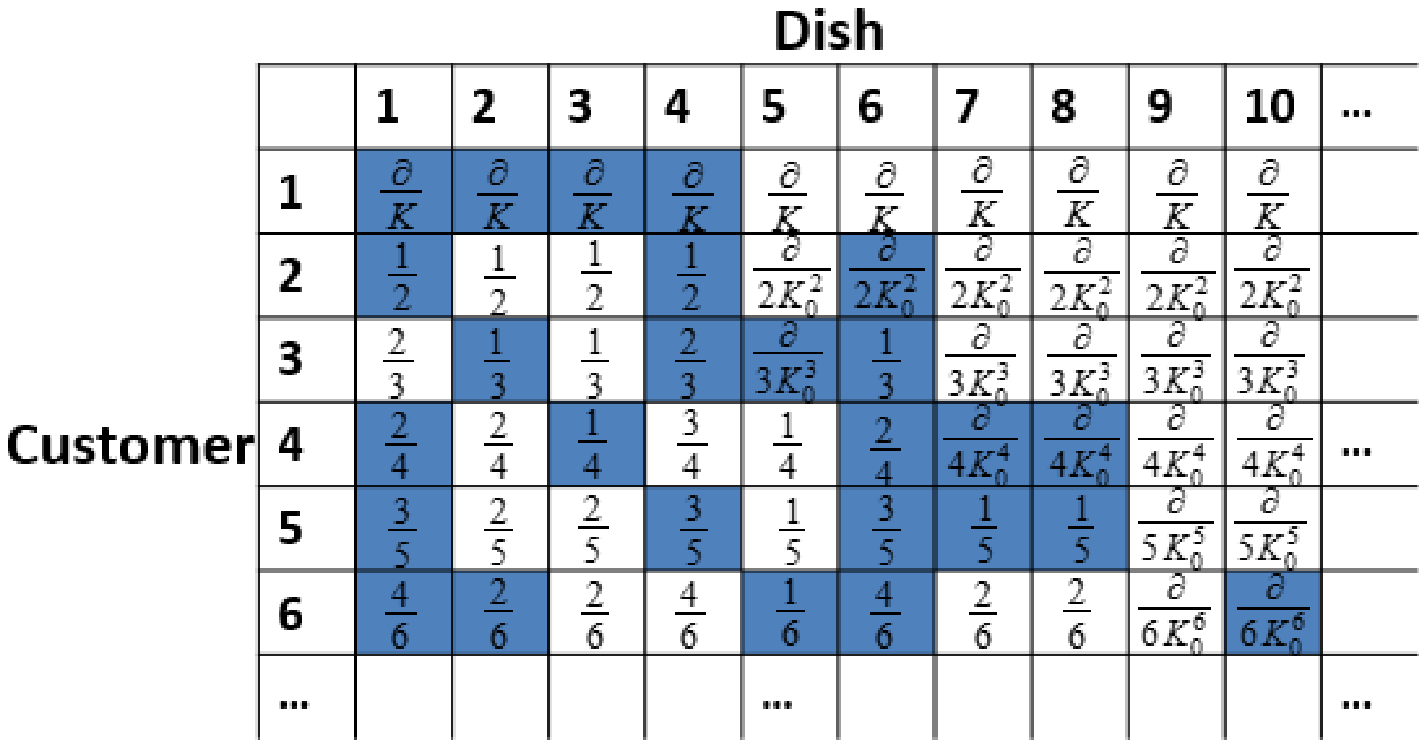}
    \vspace{-0.22cm}
    \caption{\label{fig:IBP}One realization of Indian Buffet Process.}
  \end{center}\vspace{-0.87cm}
\end{figure}
In Fig. \ref{fig:IBP}, we show one realization of an IBP. Customers are labeled by ascending numbers in a sequence. The shaded block represents the $n$th user selected dish $k$. In IBP, the first customer selects each dish with equal probability of $\frac{\alpha}{K}$, and ends up with the number of dishes following $Poisson(\alpha)$ distribution. For subsequent customers $n = 2,\ldots,N$, the probability of also having dish $k$ already belonging to previous customers is $ \frac{m_k^{n-1}}{n}$, where $m_k^{n-1}$ is the number of customers prior to $n$ with dish $k$ \cite{Griffiths.ML 2011}. Repeating the same argument as the first customer, customer $n$ will also have $m_n^0$ new dishes not tasted by the previous customers following a $Poisson(\frac{\alpha}{n})$ distribution. The probabilities of selecting certain dishes act as the prior information $\pi_k(m_k^{n-1})$. For ``old" dishes which have been tasted before, $m_k^{n-1}\neq 0$. For ``new"’ dishes which have not been sampled before, $m_k^{n-1}= 0$. After user $n$ completes its selection, $\pi_k$ will be updated to $\pi_k(m_k^{n})$. This learning process is also illustrated in Fig. \ref{fig:IBP}. $K_0^n$ is the number of dishes that have not been sampled before user $n$'s selection session.
\vspace{-0.2cm}
\section{Proposed Traffic Offloading Algorithm}
\vspace{-0.2cm}
In the previous section, we have introduced the basic model to formulate the subnetwork for D2D communication and predict users' selection. In this section, we can integrate the two layer networks together and propose the traffic offloading algorithm based on the OffSN and OnSN models.
\vspace{-0.55cm}
\subsection{System Utility}
\vspace{-0.2cm}
As the inter-OffSN interference of D2D communication can be restricted by methods such as power control \cite{Xu.2012}\cite{Golrezaei.ICC2012}, we can ignore the interference among different OffSNs. Thus, we place an emphasis on the intra-OffSN interference due to resource sharing between D2D and cellular communication. During the downlink period of D2D communication, UEs will experience interference from other cellular and D2D communications as they share the same subchannels. Thus, we can define the transmission rate of users served by the eNB as $R_c$, and by D2D communication with co-channel interference as $R_d$, \cite{Xu.2012}:
\begin{equation}\vspace{-0.2cm}
R_c= \log_2 \left(1+\frac {P_Bh_{Bc}^2}{\sum_d\beta_{cd}P_dh_{dc}^2+N_0}\right),
\end{equation}\vspace{-0.1cm}
\begin{equation}\vspace{-0.12cm}
R_d= \log_2 \left(1+\frac {P_dh_{dd}^2}{P_Bh_{Bd}^2+\sum_{d'}\beta_{dd'}P_{d'}h_{d'd}^2+N_0}\right),
\end{equation}\vspace{-0.05cm}
where $P_B$, $P_d$ and $P_{d'}$ are the transmit power of eNB, D2D transmitter $d$ and $d'$, respectively, $h_{ij}$ is the channel response of the link between UE or eNB $i$ and UE $j$, $N_0$ is the additive white Gaussian noise (AWGN) at the receivers, and $\beta_{cd}$ represents the presence of interference from D2D to cellular communication, satisfying $\beta_{cd} = 1$, otherwise $\beta_{cd} = 0$. Here, $d\neq d'$, so $\sum_{d'}\beta_{dd'}P_{d'}h_{d'd}^2$ represents the interference from the other D2D pairs that share spectrum resources with pair $d$.

The transmission rate of the users that are only served by the eNB without underlying D2D, thus also without co-channel interference is given by:
\begin{equation}\vspace{-0.25cm}
V_c= \log_2 \left(1+\frac {P_Bh_{Bc}^2}{N_0}\right).
\end{equation}\vspace{-0.25cm}

In the proposed model, even though eNB can offload traffic by D2D communication, controlling the switching over cellular and D2D communication causes extra data transmission. Thus, there exists a certain cost such as control signals transmission and information feedback during the access process \cite{Xu.2012}. Therefore, for the eNB that is serving a certain user $n$, we propose the following utility function:
\begin{equation}\vspace{-0.15cm}
U_B(n)=\left[\sum_{k=1}^{K_h}\frac {m_k^{n-1}}{n} \right]R_d+m_n^0 R_c - m_n C_c,
\end{equation}\vspace{-0.05cm}
where $C_c$ is the overhead cost for controlling the resource allocation process.
\vspace{-0.2cm}
\subsection{Proposed Algorithm}
\vspace{-0.2cm}
We propose a novel and robust algorithm that can offload the traffic of eNB without any sacrifice on the quality of service. The algorithm consists of multiple stages. In the first stage, the eNB collects the encounter history between users to compute the closeness $w_{i,j}$. Then, based on specific situation such as time and locations, eNB chooses $w_T$ dynamically. By checking if $w_{i,j}\geq w_T$, i.e., if UE $i$ and UE $j$ satisfy the predefined closeness threshold, the eNB can decide on whether to add this user into the OffSN or not. By choosing a proper $w_T$ and power control, the interference among different OffSNs can be avoided. This process will continue until no more users can be further added to the eNB's list. Then, the users in the established OffSN can construct a communication session with only intra-OffSN interference.

For websites that provide a portal to access content, such as Facebook and YouTube, the eNB will assign a special tag. Once a user visits such tagged websites, the eNB will inspect whether the user is located in an OffSN or a ``white" area. If the user is in a ``white" area, any requests of users will be served by the eNB directly. If the user is located in an OffSN, the eNB will wait the user's future activities. By browsing online, current user can have the prior information $\pi(m_k^{n-1})$ of the content distribution in the OnSN based on previous user's requests. As soon as the user requests data, the eNB detects if there are any resources in the OffSN, and then choose to set up D2D communication or not based on the feedback. For old content, the eNB will send control signal to the UE $i$ that has the highest closeness $w_{n,i}$ with user $n$. Then, UE $n$ and UE $i$ establish a D2D communication link. Even if the D2D communication is setup successfully, the eNB still waits until the data transmitting process finishes. If the D2D communication fails, the eNB will revert back to serve the user directly. For new contents, the eNB serves the user directly. After the selection is complete, the prior updates to the posterior $\pi(m_k^n)$. The proposed D2D communication algorithm is summarized in Algorithm $1$.
\vspace{-0.25cm}
\begin{algorithm}
\caption {Proposed Traffic Offloading Algorithm}
\SetAlgoLined
\KwData {$X_n$, $N_{i,j}$, $P_B$, $P_d$, $P_{d'}$}
\KwResult{$U_u(n)$, $U_B(n)$}
\textbf{1. OffSN and OnSN Generation}\;
eNB collects encounter information in cellular network\;
Find the closeness $w_{i,j}$ between two UEs\;
\If {$w_{i,j}\geq w_T$}{
Add UE $i$ and UE $j$ into OffSN\;
}
Forming OnSN by the users of corresponding UEs\;
\textbf{2. User Activity Detection}\;
\While {user $n$ is accessing the tagged websites}{
eNB detects user's location\;
 \If {user $\in$ ``white" areas}{
  eNB serves all requests directly\;
 }
 \If{user $\in$ an OffSN}{
 eNB keep watching user's activities\;
 }
}
\textbf{3. Service based on OnSN Activities}\;
\While{User $n$ is selecting contents online}{
eNB maintains prior information $\pi(m_k^{n-1})$\;
\If {old content}{
eNB locates the content holder with highest closeness $w_{n,i}$ to user $n$ in OffSN\;
Establish D2D communication\;
 \If {communication failed}{
 eNB reverts back to continue the transmission\;
 }
}
\If {new content}{
 eNB serves the request directly\;
 }
update $\pi(m_k^{n})$\;
}
\vspace{-0.15cm}
\end{algorithm}
\vspace{-0.45cm}

\section{Performance Evaluation}
\vspace{-0.3cm}
To evaluate the traffic offloading performance of our algorithm, we derive a bound on the amount of traffic that can be offloaded. We show that this problem is equivalent to the amount of contents that have been downloaded. While those locally accessible contents is related to the number of total contents and new contents selected by the users. Before we start to derive a closed-form expression on the number of old content, we first try to find its bound. Here, we adopt the Chernoff bound for analysis.
\vspace{-0.2cm}
\begin{theorem}
Let $X_1, \ldots, X_n$ be a sequence of independent trials with $P(X_i) = p_i$, $X = \sum_{i=1}^n X_i$, and $\mu = E[X]$. Then:

For any $\delta>0$, there is a bound when $X\in[0,\mu]$
\begin{equation}\vspace{-0.3cm}
P\{X<(1-\delta)\mu\}<\left[\frac{e^{-\delta}}{(1-\delta)^{(1-\delta)}}\right]^\mu,
\end{equation}\vspace{-0.05cm}
and a bound when $X\in[\mu, \infty)$
\begin{equation}\vspace{-0.3cm}
P\{X>(1+\delta)\mu\}<\left[\frac{e^{\delta}}{(1+\delta)^{(1+\delta)}}\right]^\mu.
\end{equation}\vspace{-0.05cm}
\end{theorem}

For the case of our model, the total number of contents user selects is $m_n\sim Poisson(\alpha)$, the number of new contents $m_n^0 \sim Poisson(\frac{\alpha}{n})$. Then, the number of old contents $m_n^h \triangleq m_n-m_n^0$. Hence, define the expected number of old contents $\mu \triangleq E\{m_n^h\}=E\{m_n\}-E\{m_n^0\}=\frac{n-1}{n}\alpha$, $\forall \delta > 0$, with a Chernoff bound of
\begin{equation}\vspace{-0.3cm}
\begin{split}
&P\left\{m_n^h<(1-\delta)\frac{n-1}{n}\alpha\right\}<\left[\frac{e^{-\delta}}{(1-\delta)^{(1-\delta)}}\right]^{\frac{n-1}{n}\alpha},\\
\end{split}
\end{equation}\vspace{-0.05cm}
when $m_n\in[0,\mu]$, and
\begin{equation}\vspace{-0.3cm}
\begin{split}
&P\left\{m_n^h>(1+\delta)\frac{n-1}{n}\alpha\right\}<\left[\frac{e^{\delta}}{(1+\delta)^{(1+\delta)}}\right]^{\frac{n-1}{n}\alpha},\\
\end{split}
\end{equation}\vspace{-0.05cm}
when $m_n\in[\mu,\infty)$.

As we can see, the number of old contents is the difference of two Possion distribution which follows the Skellam distribution. We can approximate pmf and CDF for the Skellam distribution using the Saddlepoint approximation. Then, we have the approximated pmf and CDF for the number of old contents as follows:
\begin{equation}\vspace{-0.2cm}
\hat{f}(k)= \frac{1}{\sqrt{2\pi (C+D)}} e^{-(\mu_{1}+\mu_{2})+C+D} \left(\frac{D}{\mu_{2}}\right)^{k},
\end{equation}\vspace{-0.05cm}
\begin{equation}\vspace{-0.2cm}
\begin{split}
\hat{F}(M)&= P\{m_n^h \leqslant M\}  \\
&= \sum_{k=0}^{M} \left[\frac{1}{\sqrt{2\pi (C+D)}} e^{-(\mu_{1}+\mu_{2})+C+D} \left(\frac{D}{\mu_{2}}\right)^{k}\right],
\end{split}
\end{equation}\vspace{-0.05cm}
where $C \triangleq \frac{k+\sqrt{k^{2}+4\mu_{1}\mu_{2}}}{2}$ and $D \triangleq \frac{2\mu_{1}\mu_{2}}{k+\sqrt{k^{2}+4\mu_{1}\mu_{2}}}$.

With the CDF function we can get the approximate number of old contents that each user select. Thus, we can estimate the traffic that can be offloaded.
\vspace{-0.15cm}
\section{Simulation Results and Analysis}
\vspace{-0.1cm}
To evaluate the performance of our algorithm, we exploit a data set of sensor mote encounter records and corresponding social network data of a group of participants at University of St Andrews by the CRAWDAD team \cite{Bigwood. WiMob2008}. In the first data set, they deployed $27$ T-mote invent devices over a period of $79$ days among $27$ users in the Department of Computer Science building. This data set helps us to establish our physical layer OffSN. In the second data set, they collected the participants' Facebook friend lists to generate a social network topology. With those information we can generate the corresponding OnSN. Then, we adopt the IBP to generate users' selections online under the assumption that the size of content library is unbounded. We assume that the content selection process has already been performed for a number of times. Thus, the eNB can obtain the prior information of the content distribution.
\begin{figure*}[htb]
\vspace{-0.7cm}
 \begin{subfigure}[b]{0.3\textwidth}
 \centering
 \includegraphics[width=\columnwidth,height=0.71\textwidth]{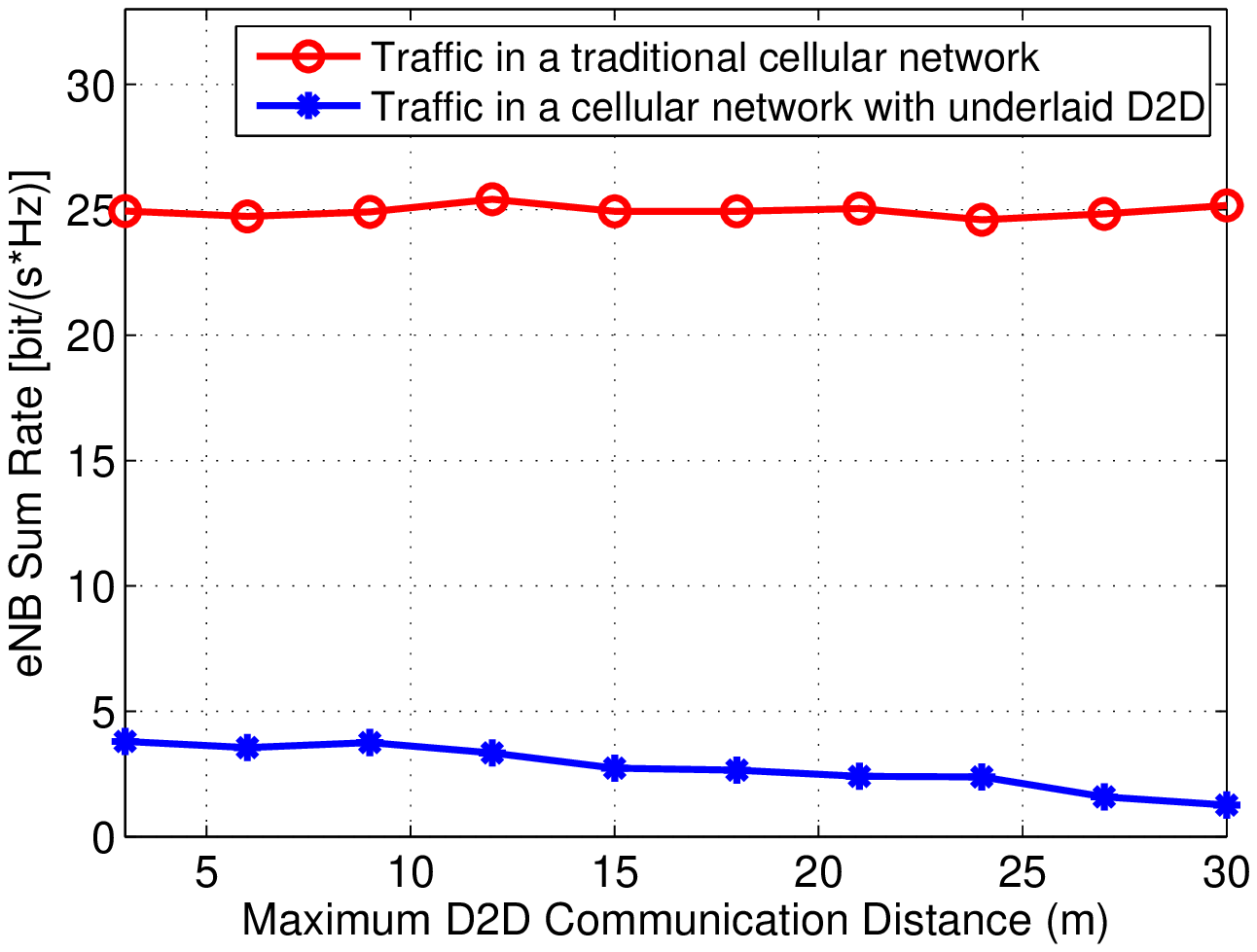}
 \caption{Average sum-rate at the eNB, as the D2D communication distance varies.}
 \label{fig:Traffic and Distance}
 \end{subfigure}
 \begin{subfigure}[b]{0.3\textwidth}
 \centering
 \includegraphics[width=\columnwidth,height=0.71\textwidth]{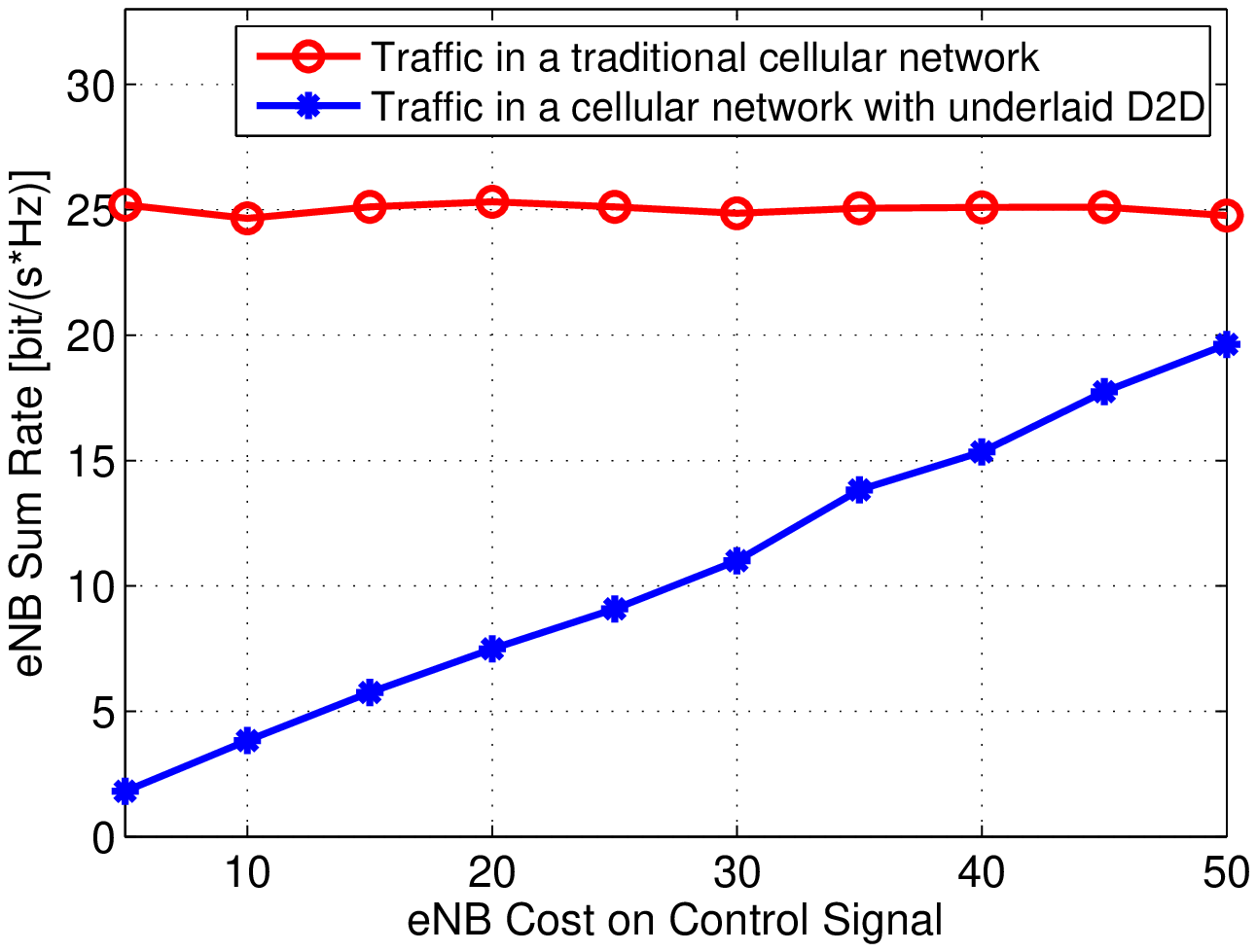}
 \caption{Average sum-rate at the eNB, as the cost for control signaling varies.}
 \label{fig:Traffic and eNB Cost}
 \end{subfigure}
  \begin{subfigure}[b]{0.3\textwidth}
 \centering
 \includegraphics[width=\columnwidth,height=0.71\textwidth]{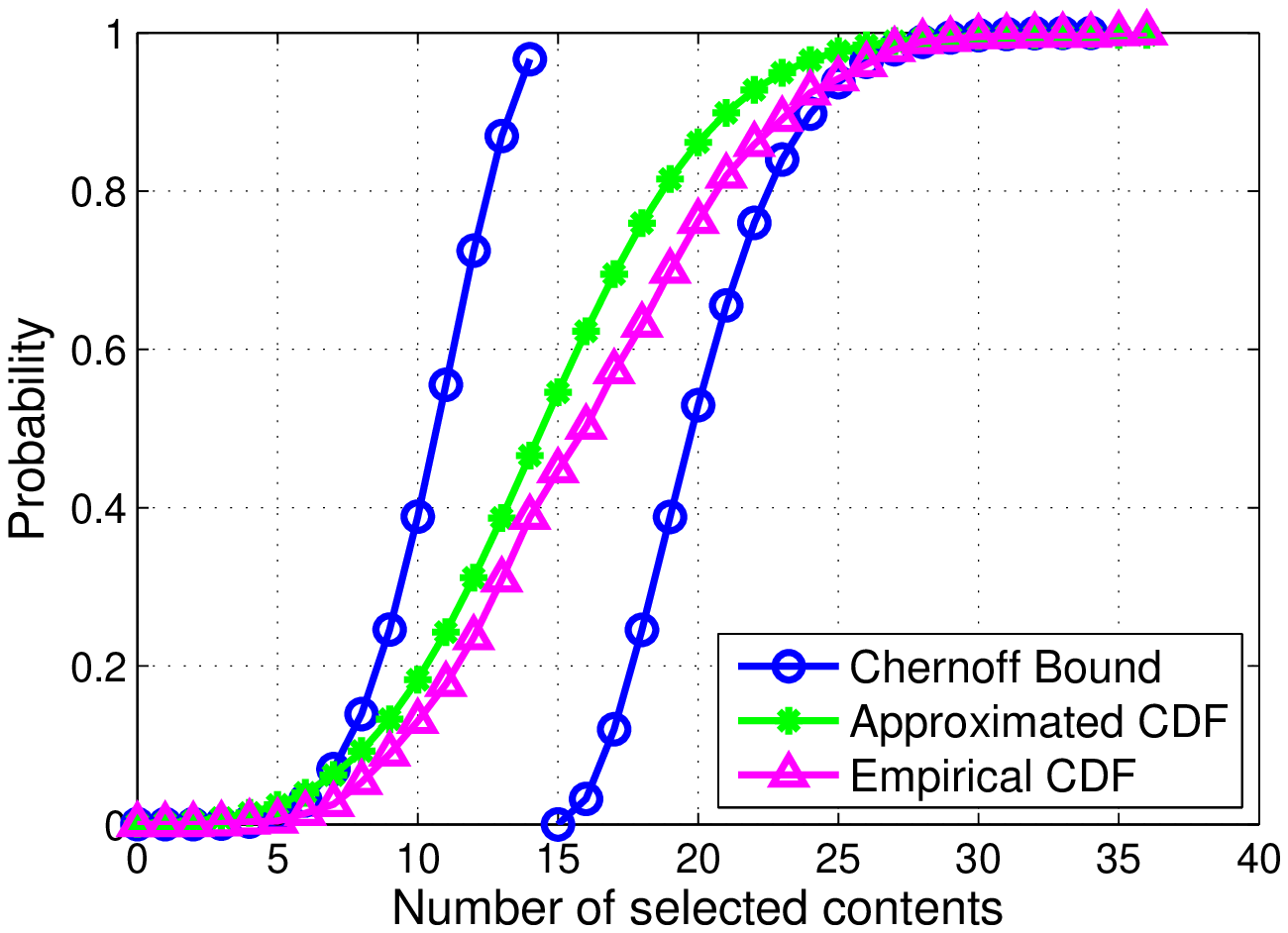}
 \caption{The Chernoff bound, approximated CDF and empirical CDF.}
 \label{fig:bound}
 \end{subfigure}
\centering
\caption{Simulation results.}
\label{fig:SIM}
\vspace{-0.7cm}
\end{figure*}
As the D2D communication distance increases, the eNB will have more possibilities for detecting available contents providers. As a result, the performance of traffic offloading will be better with a larger maximum distance. This assumption is shown in Fig. \ref{fig:Traffic and Distance}. In this figure, we can see that, increasing the maximum communication distance, yields a decrease in the eNB's data rate and an increase in the amount of offloaded traffic. However, we note that, with the increase of the transmission distance, the associated UE costs (e.g., power consumption) will also increase. Thus, the increase of D2D communication distance will provide additional benefits to the eNB, but not for users.

In Fig. \ref{fig:Traffic and eNB Cost}, we show the variation of the sum-rate at the eNB as the cost for control signal varies. As the eNB has to arrange the inter change process between cellular and D2D communication, necessary control information is needed. Moreover, additional feedback signals are required for monitoring the D2D communication and checking its status. Those costs will affect the traffic offloading performance of the system. In our simulation, we define the cost as the counteract to the gain in data rate from 5\% to 50\%.  As we can see, increasing the cost on control signal, the offload traffic amount is decreased.

In order to know the amount of traffic that can be offloaded, we specified a special case when $\alpha=20$, then plot the Chernoff bound and Saddlepoint approximation of the CDF of the $4$th user's number of old contents in Fig. \ref{fig:bound}. As we can see, the approximated CDF lies between the Chernoff bound. We have mentioned in the previous section, the mean value of the number of old contents is $E\{m_n^h\}=\frac{n-1}{n}\alpha=15$. So there is a gap between the number of $14$ and $15$. Then we simulate the $4$th user's selection and plot the empirical CDF. The simulated empirical CDF also lies between the upper and lower Chernoff bound as we expected. In addition, The approximated CDF line is quite close to the simulated empirical CDF line, which proves our analysis in the previous section.

\vspace{-0.2cm}
\section{Conclusions}
\vspace{-0.2cm}
In this paper, we have proposed a novel approach for improving the performance of D2D communication underlaid over a cellular system, by exploiting the social ties and influence among individuals. We formed the OffSN to divide the cellular network into several subnetworks for carrying out D2D communication with only intra-OffSN interference. Also we established the OnSN to analyse the OffSN users' online activities. By modeling the influence among users on contents selection online using the Indian Buffet Process, we have obtained the distribution of contents requests, and thus can get the probabilities of each contents to be requested. Using our proposed algorithm, the traffic of eNB has been reduced. Simulation results based on real traces have shown that different parameters for eNB and users will lead to different traffic offloading performances.

\vspace{-0.2cm}

\bibliographystyle{IEEEtran}

%


\end{document}